*Aligning Artificial Intelligence with Humans through Public Policy*

John Nay, James Daily

May 4, 2022

**Abstract**

Given that Artificial Intelligence (AI) increasingly permeates our lives, it is critical that we systematically align AI objectives with the goals and values of humans. The human-AI alignment problem stems from the impracticality of explicitly specifying the rewards that AI models should receive for all the actions they could take in all relevant states of the world. One possible solution, then, is to leverage the capabilities of AI models to learn those rewards implicitly from a rich source of data describing human values in a wide range of contexts. The democratic policy-making process produces just such data by developing specific rules, flexible standards, interpretable guidelines, and generalizable precedents that synthesize citizens' preferences over potential actions taken in many states of the world. Therefore, computationally encoding public policies to make them legible to AI systems should be an important part of a socio-technical approach to the broader human-AI alignment puzzle. Legal scholars are exploring AI, but most research has focused on how AI systems fit within existing law, rather than how AI may understand the law. This Essay outlines research on AI that learn structures in policy data that can be leveraged for downstream tasks. As a demonstration of the ability of AI to comprehend policy, we provide a case study of an AI system that predicts the relevance of proposed legislation to a given publicly traded company and its likely effect on that company. We believe this represents the "comprehension" phase of AI and policy, but leveraging policy as a key source of human values to align AI requires "understanding" policy. We outline what we believe will be required to move toward that, and two example research projects in that direction. Solving the alignment problem is crucial to ensuring that AI is beneficial both individually (to the person or group deploying the AI) and socially. As AI systems are given increasing responsibility in high-stakes contexts, integrating democratically-determined policy into those systems could align their behavior with human goals in a way that is responsive to a constantly evolving society.

## I. Introduction

Artificial Intelligence (AI) permeates our lives.[1] Large neural network models have exceeded human-level performance on difficult tasks[2] and learned useful knowledge without explicit supervision related to that knowledge,[3] which increases the probability that AI will be entrusted with a broader range of responsibilities.[4] As models scale in size, compute, and data inputs, there may be emergent breakthroughs in their capabilities.[5] "Foundation models" are a new class of deep learning AI models that are effective across multiple tasks and are poised for widespread deployment, despite the research community having an incomplete understanding of their capabilities and shortcomings.[6] Given that we are increasingly ceding human activities to AI,[7] we should attempt to systematically align AI objectives with the goals and values of the humans they serve.[8]

AI models often "hack" the reward systems used for their training by coming up with unanticipated "shortcut" behaviors that optimize for the explicit (inherently limited) objectives set out by the human designers but do not accomplish the ultimately intended outcome.[9] Recent empirical results suggest that this problem is likely to worsen as models become more flexible and powerful.[10]

---

[1] For example, AI increasingly determines the media we see (social media feeds, content recommendation systems) how we travel (vehicle automation), and our health (drug discovery, radiology interpretation). *See*, *e.g.*, Sang Ah Kim, *Social Media Algorithms: Why You See What You See*, 2 GEO. L. TECH. REV. 147 (2017); Silvia Milano et al., *Recommender Systems and Their Ethical Challenges*, 35 AI & SOC'Y 957 (2020); National Conference of State Legislatures, *Autonomous Vehicles | Self-Driving Vehicles Enacted Legislation*, ncsl.org (Feb. 18, 2020), https://www.ncsl.org/research/transportation/autonomous-vehicles-self-driving-vehicles-enacted-legislation.aspx; Bibb Allen et al., *2020 ACR Data Science Institute Artificial Intelligence Survey*, 18 J. Am. Col. Radiology 1153, 1156 (2020) ("[a]pproximately 30% of radiologists are currently using AI as part of their practice" in 2020); Neil Savage, *Tapping Into the Drug Discovery Potential of AI*, BIOPHARMA DEALMAKERS, May 27, 2021, https://www.nature.com/articles/d43747-021-00045-7 (discussing the "many companies founded in the past decade around AI-based strategies for drug discovery and development").
[2] Andrew W. Senior et al., *Improved Protein Structure Prediction Using Potentials From Deep Learning*, 577 NATURE 706 (2020) (describing AlphaFold, a major improvement in the problem of predicting the structure of a protein from its genetic sequence); Aditya Ramesh et al., *Hierarchical Text-Conditional Image Generation with CLIP Latents*, arxiv.org (Apr. 13, 2022), https://arxiv.org/abs/2204.06125 (describing DALL•E 2, an AI system that can create human-quality images from natural language descriptions).
[3] *See, e.g.,* Aakanksha Chowdery et al., *PaLM: Scaling Language Modeling with Pathways*, arxiv.org (Apr. 7, 2022), https://arxiv.org/pdf/2204.02311.pdf (exceeding average human performance on a set of 58 complex natural language tasks from the Beyond the Imitation Game Benchmark, including guessing which English language proverb best describes a passage of text, explaining the result of a series of navigational instructions, and putting a series of months, actions, numbers, etc. into a logical ordering).
[4] *On the current state of AI and its deployment, see* Daniel Zhang, et al., *The AI Index 2022 Annual Report*, Stanford Institute for Human-Centered Artificial Intelligence (March 2022), https://aiindex.stanford.edu/wp-content/uploads/2022/03/2022-AI-Index-Report_Master.pdf.
[5] Katja Grace et al., *When Will AI Exceed Human Performance? Evidence from AI Experts*, 62 J. ARTIFICIAL INTEL. RSCH. 729 (2018).
[6] Rishi Bommasani et al., *On the Opportunities and Risks of Foundation Models*, arxiv.org (Aug. 18, 2021) https://arxiv.org/pdf/2108.07258.pdf.
[7] *For more on the purpose (generalizability) of real-world AI systems and the division of tasks between humans and AI, see,* John Nay & Katherine Strandburg, *Generalizability: Machine Learning and Humans-in-the-loop*, in BIG DATA LAW (Roland Vogl ed., 2021).
[8] BRIAN CHRISTIAN, THE ALIGNMENT PROBLEM: MACHINE LEARNING AND HUMAN VALUES (2020); Iason Gabriel, *Artificial Intelligence, Values, and Alignment*, 30 MINDS & MACHINES 411 (2020); STUART RUSSELL, HUMAN COMPATIBLE: ARTIFICIAL INTELLIGENCE AND THE PROBLEM OF CONTROL (2019).

However, how best to pursue AI alignment is not yet clear.[11] The U.S. National Institute of Standards and Technology (NIST) final report on identifying and managing bias in AI argues that "what is missing from current remedies is guidance from a broader socio-technical perspective that connects these practices to societal values."[12] Such a perspective "considers AI within the larger social system in which it operates" and "take[s] into account the needs of individuals, groups and society."[13] Stefano Ermon, Associate Professor of Computer Science at Stanford University, has argued that human-AI alignment "is something that I think the majority of people would agree on, but the issue, of course, is to define what exactly these values are."[14] A recent paper from a large AI research company, OpenAI, concluded that "one of the biggest open questions is how to design an alignment process that is transparent, that meaningfully represents the people impacted by the technology, and that synthesizes peoples' values in a way that achieves broad consensus amongst many groups."[15]

In a democracy, policy-making is a continuous — and ideally transparent and representational — effort to "synthesize peoples' values in a way that achieves broad consensus amongst many groups" as an expression of societal values. Therefore, computationally encoding proposed and enacted public policies to make them legible to AI systems should be an important part of a socio-technical approach to the broader human-AI alignment puzzle.[16]

## II. Alignment with Policy through Machine Learning

Aligning AI with public policy will require consideration of as much data as possible in order to have a comprehensive view with sufficient depth. The U.S. federal government alone produces an enormous amount of policy data, on the order of a million pages of bills and regulations every year.[17] Learning structure in near real-time from source material at that scale is more plausibly tractable through a machine learning (ML) approach that can automatically incorporate new information, relative to a human-led approach of "hard-coding" and "hand-engineering" data into computational representations. Recent breakthroughs in AI have occurred through applying ML to massive amounts of data with little human supervision or data labeling. For policy-AI alignment to keep pace with broader ML developments, we should take a similar approach as a component of the toolkit.[18]

---

[9] *See, e.g.,* W. Bradley Knox et al., *Reward (Mis)design for Autonomous Driving*, arxiv.org (Mar. 11, 2022), https://arxiv.org/abs/2104.13906; Victoria Krakovna et al., *Specification Gaming: The Flip Side of AI Ingenuity*, deepmind.com (Apr. 21, 2020), https://www.deepmind.com/blog/specification-gaming-the-flip-side-of-ai-ingenuity.
[10] Alexandra Pan et al., *The Effects of Reward Misspecification: Mapping and Mitigating Misaligned Models*, arxiv.org (Feb. 14, 2022), https://arxiv.org/abs/2201.03544.
[11] Reva Schwartz et al., *Towards a Standard for Identifying and Managing Bias in Artificial Intelligence*, NIST Special Publication 1270 at 10 (Mar. 2022), https://doi.org/10.6028/NIST.SP.1270 ("While ML systems are able to model complex phenomena, whether they are capable of learning and operating in line with our societal values remains an area of considerable research and concern").
[12] Schwartz et al. at ii.
[13] *Id*. at 10-11.
[14] Ariel Conn, *Stefan Ermon Interview*, Future of Life Institute (Jan. 26, 2017), https://futureoflife.org/2017/01/26/stefano-ermon-interview/.
[15] Long Ouyang et al. *Training Language Models to Follow Instructions with Human Feedback* 20, arxiv.org (Mar. 4, 2022), https://arxiv.org/pdf/2203.02155.pdf
[16] There are, of course, many important human values not encoded into public policy, and democratic policy-making is not a perfect process; however it may be the best approach available. *See* WINSTON CHURCHILL, WINSTON S. CHURCHILL: HIS COMPLETE SPEECHES, 1897-1963 at 7566 (Robert Rhodes ed. 1974).
[17] *See*, *e.g.*, congress.gov (20,253 bills introduced in the 116th Congress); federalregister.gov (27,712 documents published in 2021).

Advancements in natural language processing (NLP) have been driven primarily by scaling the training of flexible neural-network-based language model architectures to enormous amounts of text.[19] These large ML models are trained on relatively simple tasks but can learn complex semantic relationships, even between words and phrases that do not appear similar based on their characters.[20] For instance, a common method is to teach the model to predict the missing word in a sentence (e.g. "___ grow on trees."). Because the model has been fed billions of words of text, including numerous books and websites about horticulture, it knows that words such as apples, pears, and oranges are likely to be the missing word. As a result, the model learns that these words are semantically related even though they are orthographically different. From additional examples the model learns that the words apples and pears are more closely related than the words apples and oranges. Surprisingly, initial training on basic tasks such as this allows the resulting models to serve as core building blocks for more specialized tasks without specific additional downstream training (so-called "zero shot" learning).[21]

If the process of "pre-training" a foundation model on a self-supervised task (e.g., predicting held-out words from a sentence, or whether one sentences is next to another), and then porting generalizable baseline capabilities to a different task (e.g., question answering) is a useful framework for building capable automated systems,[22] then training ML models on tasks that require learning latent structure of law and policy could be an important step toward embedding more generalizable knowledge of policy goals in advanced AI deployed on downstream end-tasks that are not necessarily directly related to policy.

At this early phase of AI alignment, we are focused on ML models that learn structures in policy data that can be leveraged for a relatively limited set of downstream tasks. As a stepping stone toward more ambitious alignment, we are aiming for methods that make policy legible to AI.[23] We call the near-term goal "policy comprehension" and the long-term goal "policy understanding." We define "comprehension" as sufficient performance on a task of the same nature as the training task but with data the model has not seen during training, whereas "understanding" is the ability to generalize from what has been comprehended into sufficient performance on novel downstream tasks with minimal additional training for those tasks. "Sufficient performance" simply means good enough to deploy the resulting system in a real-world application. For a task that is currently economical for humans to perform, deployment may require human-or-better performance, but lower levels of performance can be sufficient for tasks that are beyond human scale.

---

[18] For a discussion of regulations themselves keeping up with quickly advancing technology such as AI, *see* Mark D. Fenwick et al., *Regulation Tomorrow: What Happens When Technology Is Faster than the Law?*, 6 AM. U. BUS. L. REV. 561 (2017).

[19] LEWIS TUNSTALL ET AL., NATURAL LANGUAGE PROCESSING WITH TRANSFORMERS, xii (2022) ("In just a few years [the transformer neural network architecture] swept across the field, crushing previous architectures …[It] is excellent at … dealing with huge datasets."); Transformer deep neural networks can capture medium-range dependencies, and interactions, within natural language. Ashish Vaswani et al., *Attention Is All You Need*, *in* PROCEEDINGS OF THE 31ST CONFERENCE ON NEURAL INFORMATION PROCESSING SYSTEMS (2017).

[20] "[...] semantic knowledge about context-dependent similarities is explicitly represented in the structure of word embeddings. Thus, extremely detailed conceptual knowledge can be constructed bottom up by merely tracking word co-occurrence statistics." Gabriel Grand et al., *Semantic Projection Recovers Rich Human Knowledge of Multiple Object Features from Word Embeddings, Nature Human Behavior* (Apr. 14, 2022), https://www.nature.com/articles/s41562-022-01316-8.

[21] *See, e.g.,* Jason Wei et al., *Finetuned Language Models Are Zero-Shot Learners*, arxiv.org, (Feb. 8, 2022), https://arxiv.org/abs/2109.01652.

[22] *See, e.g.,* Drew A Hudson and C. Lawrence Zitnick, *Generative Adversarial Transformers*, *in* PROCEEDINGS OF THE 38TH INTERNATIONAL CONFERENCE ON MACHINE LEARNING (Marina Meila & Tong Zhang eds. 2021).

We use the terms comprehension and understanding from a pragmatist perspective of the philosophy of language. This perspective is concerned only with the model's empirically observable behavior and not with its internal structure or standalone ability to evaluate the truth of statements in relation to the physical world.[24] From this perspective, comprehension and understanding are judged by the model's performance in the task at hand (i.e., "meaning as use," in which the meaning of words is derived from the results of their use in a particular context).[25] We explicitly do not concern ourselves with whether a model understands language in the same way that a human does.[26]

In the legal context, an example of comprehension would be a model trained on identifying non-standard contract clauses accurately classifying a contract clause that the model was not trained on. Understanding goes a step further. For example, a model trained only on pre-GDPR law correctly determining whether a privacy policy complies with the GDPR given only the text of the GDPR and the privacy policy would qualify as understanding.[27] A particular challenge the legal domain poses to AI is to correctly reason about a large set of novel facts under an ever-changing body of rules.[28]

### III. A Case Study of AI Policy Comprehension

As a demonstration of the ability of AI to comprehend policy, we provide a case study of an AI system that predicts the relevance of proposed legislation to a given publicly traded company and its likely effect on that company. A significant problem with many tests of AI systems is that they are conducted on "toy" problems or on historical data that is overfit.[29] In order to assess policy comprehension, we tested AI predictions against financial market responses to anticipated policy events in real-time. This avoids both the look-ahead bias that plagues most analyses (especially those using financial market data[30]) and the toy-problem nature of AI applications in games or other purely simulated scenarios.

---

[23] *See, e.g.,* Daniel Martin Katz & John Nay, *Machine Learning and Law*, *in* LEGAL INFORMATICS (Daniel Martin Katz et al. eds., 2021).
[24] Bommasani et al., *supra* at 50.
[25] LUDWIG WITTGENSTEIN, PHILOSOPHICAL INVESTIGATIONS § 43 (John Wiley & Sons 2009).
[26] *Cf* Chris Potts, *Is it Possible for Language Models to Achieve Language Understanding?*, medium.com (Oct. 5, 2020), https://chrisgpotts.medium.com/is-it-possible-for-language-models-to-achieve-language-understanding-81df45082ee2 (defining "understanding in terms of robustly acquiring very complex language capabilities, rather than in terms of human-like capacities or behaviors").
[27] The European Union's General Data Protection Regulation (GDPR) required or incentivized many companies worldwide to adopt more stringent privacy policies in order to comply with the GDPR. *See, e.g.,* Michael L. Rustad & Thomas H. Koenig, *Towards a Global Data Privacy Standard*, 71 FLA. L. REV. 365, 387-395 (2019).
[28] Bommasani et al., *supra* at 15-16.
[29] MARCOS LÓPEZ DE PRADO, ADVANCES IN FINANCIAL MACHINE LEARNING 151-54 (2018) ("Most backtests published in journals are flawed, as the result of selection bias on multiple tests").

### A. Overview of the AI Models

Every day, for each U.S. public company, the AI:
1. Identifies pending legislation that is likely to be relevant to that company;
2. Predicts whether the legislation would likely have a positive or negative impact on the company;
3. Predicts the overall significance of the legislation to companies;
4. And predicts the probability that the bill will be enacted into law.

Because almost all law is expressed in natural language, NLP is the foundation of this system. The text of each bill is parsed and converted into mathematical representations legible to ML using language models (of the type described above). Similar pipelines are required for each of the several streams of data used to characterize each company for purposes of identifying relevant legislation and predicting its effects. The four component ML models are trained, and validated out-of-sample, using more than one million labels. The result is a system that accurately predicts the enactment likelihood and significance of a bill, draws a connection of relevance between even a small subsection of that bill and a single part of a company's business model, and predicts its likely effect on that company.

We then categorize bills as either Democratic or Republican using the party affiliation and partisanship of each bill's Congressional sponsor and cosponsors (if any). This aggregation groups the AI's company-impact predictions by political party and therefore produces an electoral impact model for validation against financial markets.

### B. Empirical Validation

Deeply liquid markets are efficient information processors. The prices that emerge encapsulate the "wisdom of the crowd" and its views on events expected to impact future value, which provides a source of data for assessing whether the AI captured the intended policy-related phenomena.

Some companies are more likely to be positively (or negatively) affected by a political party's policies. As an election approaches, companies that would be positively affected by the expected outcome are more likely to increase in price compared to those that would be negatively affected by that outcome. If the AI predictions of the potential impacts of a party's proposed policies on individual companies are accurate, the spread between the returns of long positions (held in companies the AI predicts would be beneficiaries if that party wins) and short positions (held in companies the AI predicts would be losers if that party wins) should reflect the market's expected electoral outcome for that party. To the extent that the long-short portfolio return is orthogonal to other market factors, the changing value of the portfolio of equity shares reflects the market's changing expectation of the likelihood of the electoral outcome and the magnitude of its impact.[31]

---

[30] *Id.*

To test this, we use the AI output to estimate a political party's potential policy impact on each company and select the top and the bottom ranked companies for long and short positions. Democratic and Republican rankings were separately constructed using the same methodology, differing only in the party affiliation of the underlying bills. We rebalanced the Republican and Democrat index portfolios every two weeks to keep the implemented portfolio in line with the underlying signal of company predictions. We used point-in-time prediction data to generate 2016 indices, and simulated their investment returns with point-in-time price data (Fig. 1).

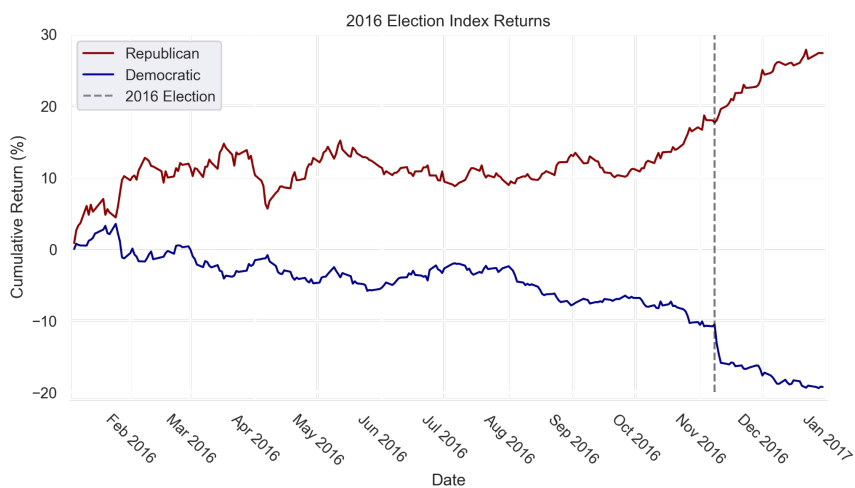

**Figure 1.** Index returns in 2016.

These results support our hypothesis that the AI captures the political exposures through the policy text data. The Republican index outperformed the Democratic index in the lead-up to and, especially, during the immediate aftermath of the election, implying a Republican win.[32] The Republicans retained control of the House and Senate, and gained control of the Presidency.

In 2020 we generated the indices in real-time, and billions of dollars were traded on financial products based on the indices. The Democratic index broke away as the leader on March 19th.[33] June 8th marked a turning point, with a clear acceleration in the Democratic index and the Republic index moving into negative territory soon after. The Democrats won the Presidency, Senate, and House.

---

[31] Note that we are referring to changes because markets rapidly incorporate new information and existing information is generally already "priced in." In this Essay, we do not address teasing apart likelihood and magnitude.

[32] *See, e.g.,* Nate Silver et al., 2016 Election Forecast, fivethirtyeight.com (Nov. 8, 2016), https://projects.fivethirtyeight.com/2016-election-forecast/.

[33] 538 did not begin making predictions until June 1st, 2020, with a Biden win probability of 70%, which did not exceed 80% until October. Nate Silver et al., *2020 Election Forecast*, fivethirtyeight.com (Nov. 3, 2020), https://projects.fivethirtyeight.com/2020-election-forecast/.

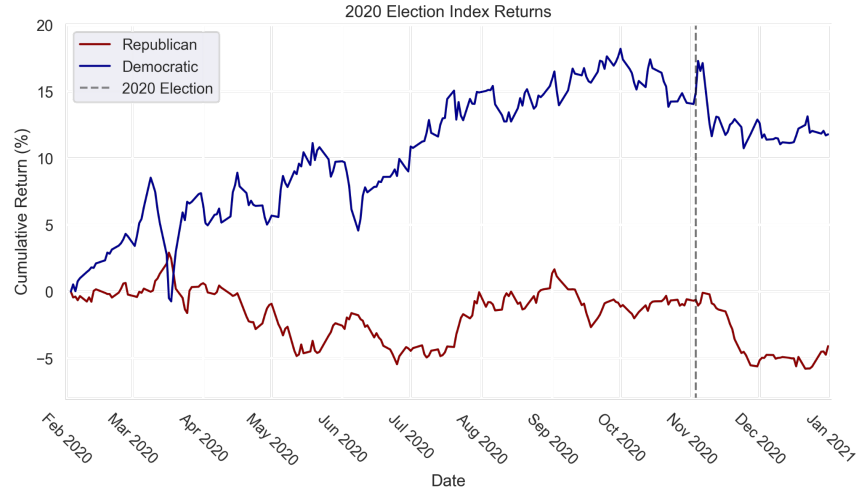

**Figure 2.** Index returns in 2020.

The results illustrate AI comprehension of policy data in a manner that was useful for a real-world task. The AI system was trained on historical data on human judgments of the relevance of companies to historically proposed policies and then deployed in a live situation where it assessed the relevance of new companies to new policies. Its predicted outputs were relied on by investment professionals. Although this represents useful comprehension of policy data for a downstream task on new data that was not used to train the AI, this does not qualify as AI understanding policy under our definition because the AI was not applied to completely novel tasks without training for those tasks.

## IV.  Toward Policy Understanding

Legal scholars are exploring AI, but most research has focused on how AI systems fit within existing law, rather than how AI may understand the law.[34] AI can currently learn basic structures in law and policy data,[35] and then leverage that learning for limited downstream tasks such as identifying companies that are relevant to policy events. We believe this represents the comprehension phase of AI and policy, but leveraging policy as a key source of human values to align AI with human values requires understanding policy.

The naïve approach of asking a foundation model general legal reasoning questions does not work (yet) for demonstrating an understanding of policy. Progress will likely require a combination of methodological (e.g., model architecture or model training procedures) and data improvements. We believe we need to construct richer data sources that more directly train models on policy structures, rather than only training on raw text with simple self-supervision. In Section IV, we outline two example research projects on this possible path.

---

[34] *See, e.g.,* Solon Barocas & Andrew D. Selbst, *Big Data's Disparate Impact*, 104 CAL. L. REV. 671 (2016); Matthew U. Scherer, *Regulating Artificial Intelligence Systems: Risks, Challenges, Competencies, and Strategies,* 29 HARV. J. L. & TECH. 353 (2016).

[35] For examples of AI learning useful structures in law and policy text data, *see, e.g.,* John Nay, *Predicting and Understanding Law-making with Word Vectors and an Ensemble Model*, 12 PLOS ONE 1 (2017); J.B. Ruhl, John Nay, Jonathan Gilligan, *Topic Modeling the President: Conventional and Computational Methods*, 86 GEO. WASH. U. L. REV. 1243 (2018); John Nay, *Gov2Vec: A Case Study in Text Model Application to Government Data*, *in* LEGAL INFORMATICS (Daniel Martin Katz et al. eds. 2021); Ilias Chalkidis et al., *LexGLUE: A Benchmark Dataset for Legal Language Understanding in English*, *in* PROCEEDINGS OF THE 60TH ANNUAL MEETING OF THE ASSOCIATION FOR COMPUTATIONAL LINGUISTICS (2022). *See, generally,* Harry Surden, *Artificial Intelligence and Law: An Overview*, 35 GA. ST. U. L. REV. 1305 (2019).

Many financial firms seek to cultivate a "culture of compliance" in order to imbue the spirit of the regulatory regime in which the firm operates into the everyday actions of its employees. Analogously, an ML-driven understanding of existing and proposed laws, regulatory standards and legal precedents could be embedded into AI models operating within highly regulated environments such as healthcare data privacy, taxation compliance, and investment advising. To continue with the financial markets case study theme, the following is an example of understanding in which training on regulatory compliance may improve holistic performance on primary tasks such as investment advising.

1. Train an ML model on question-answer pairs from regulatory exams that human financial advisors must pass to provide investment advice in the U.S. securities industry, compliance training materials, and regulatory enforcement manuals.
2. After any fine-tuning on additional training data for the downstream primary task not explicitly related to regulatory compliance (whether that be in simulated environments, with live human-in-the-loop feedback, or offline), such as investment portfolio construction, deploy the resulting model on those tasks.
3. If the initial regulatory training process was able to imbue an understanding of the policy regime, then the model deployed on its primary investing task would be in compliance more often than a model that was not subjected to the initial training.

Another example is to apply language models within the regulatory compliance function, but across regulatory areas. This could be accomplished by tuning a large language model on regulatory text and fine-tuning on "chain-of-thought" prompts from sources such as compliance training materials and exams. These prompts would be presented in an issue-rule-application-conclusion (IRAC) format familiar to law students. The resulting model could be validated by applying it to novel regulatory issues in areas on which it was not trained. The general feasibility of this style of chain-of-thought prompting of large language models was recently illustrated by Google AI's PaLM model.[36]

## V. Conclusion

We believe policy-AI alignment will be an increasingly important tool for shaping AI development. The human-AI alignment problem stems primarily from the impracticality of explicitly specifying the rewards that AI models should receive for all the actions they could take in all relevant states of the world. One possible solution, then, is to leverage the capabilities of AI models to learn those rewards implicitly from a rich source of data describing human values in a wide range of contexts. The democratic policy-making process produces just such data by developing specific rules, flexible standards, interpretable guidelines, and generalizable precedents that synthesize citizens' preferences over potential actions taken in many states of the world. The ability of AI models to comprehend and—eventually—understand complex natural language data could provide a generalizable solution to the alignment problem.

---

[36] Chowdery et al., *supra* note 1.

Solving the alignment problem is crucial to ensuring that AI is beneficial both individually (to the person or group deploying the AI) and socially. As AI systems are given increasing responsibility in high-stakes decision-making contexts, integrating democratically-determined policy into those systems will align their behavior with human goals in a way that is responsive to a constantly evolving society.

**Disclaimer**

*This Essay is not to be construed as an offer to sell or the solicitation of an offer to buy any security, nor is it an offer to act as an investment advisor. Any such offer will only be made by an express declaration accompanied by all required documentation. To the extent that performance information is contained in this communication, past performance is not necessarily indicative of the future results and all figures are estimated and unaudited unless otherwise noted.*